\documentclass[a4paper,11pt]{article}
\usepackage{pos}

\title{Lattice study of RG fixed point based on gradient flow
in $3$D $O(N)$ sigma model}

\author*[a]{Okuto Morikawa}
\author[b]{Mizuki Tanaka}
\author[c,d]{Masakiyo Kitazawa}
\author[e]{Hiroshi~Suzuki}

\affiliation[a]{Interdisciplinary Theoretical and Mathematical Sciences Program (iTHEMS),
RIKEN, Wako 351-0198, Japan}
\affiliation[b]{Department of Physics, Osaka University, Toyonaka,
Osaka 560-0043, Japan}
\affiliation[c]{Yukawa Institute for Theoretical Physics,
Kyoto University, Kyoto 606-8502, Japan}
\affiliation[d]{
  J-PARC Branch, KEK Theory Center, 
  Institute of Particle and Nuclear Studies, KEK, Tokai, Ibaraki 319-1106, Japan}
\affiliation[e]{Department of Physics, Kyushu University,
744 Motooka, Nishi-ku, Fukuoka 819-0395, Japan}

\emailAdd{okuto.morikawa@riken.jp}
\emailAdd{kitazawa@yukawa.kyoto-u.ac.jp}
\emailAdd{hsuzuki@phys.kyushu-u.ac.jp}

\abstract{
We present the lattice simulation of the renormalization group flow
in the $3$-dimensional $O(N)$ linear sigma model.
This model possesses a nontrivial infrared fixed point, called Wilson--Fisher fixed point.
Arguing that the parameter space of running coupling constants can be spanned by
expectation values of operators evolved by the gradient flow,
we exemplify a scaling behavior analysis based on the gradient flow
in the large $N$ approximation at criticality.
Then, we work out the numerical simulation of the theory with finite~$N$.
Depicting the renormalization group flow along the gradient flow,
we confirm the existence of the Wilson--Fisher fixed point non-perturbatively.
}

\FullConference{The 41st International Symposium on Lattice Field Theory (LATTICE2024)\\
 28 July - 3 August 2024\\
Liverpool, UK\\}

\begin{document}
\maketitle

\section{Introduction and summary}
The renormalization group (RG)~\cite{Wilson:1971bg,Wilson:1971dh} has been instrumental
in understanding scale-dependent phenomena and phase transitions at criticality.
A significant step of its transformation may be provided by
the coarse-graining such as spin blocking or integrating over higher momentum modes.
Similarly, in perturbative RG technique in quantum field theory,
the change of the renormalization scale gives rise
to effectively energy-dependent coupling constants,
or the running couplings being subject to an RG equation.
Outside the perturbative regime,
we expect that coupling constants flow along RG
and some of them hopefully possess critical values
at RG-invariant fixed points~\cite{Kadanoff:1966wm,Mack:1969rr,Polyakov:1970xd}.
Attractive issues in every quantum system \textit{sub specie aeternitatis},
as discussed in Ref.~\cite{Wilson:1973jj},
are concentrated on clarifying it as universal scaling law near a nontrivial fixed point.

For half a century, physicists and mathematicians have studied the RG in nature,
e.g., based on perturbative calculations, $\epsilon$ expansion,
large $N$ approximation, and so on.
See Ref.~\cite{Zinn-Justin:2002ecy} from the physical point of view.
Despite remarkable achievements,
there has been no available and effective formulation
of the non-perturbative RG in any gauge theory.
The functional definition of RG requires a momentum cutoff function
and then is usually incompatible with gauge symmetry.
For attempts to make a smooth momentum cutoff compatible with gauge symmetry,
see Refs.~\cite{Sonoda:2007dj,Igarashi:2009tj}.

To put this program into practice,
it would be a wonder~\cite{Luscher:2013vga} that the RG flow can be identified
with the so-called gradient flow~\cite{Narayanan:2006rf,Luscher:2009eq,Luscher:2010iy,Luscher:2011bx}.
The gradient flow is a kind of
diffusion equation evolving fields $\{\mathcal{A}_i\}$ along the fictitious time~$t$,
\begin{equation}
 \partial_t\mathcal{A}_i(t,x) = \partial_\mu^2\mathcal{A}_i(t,x) + \dots,
\end{equation}
the leading formal solution
$\mathcal{A}_i(t,x)\sim t^{-D/2}\int d^D y\, e^{-(x-y)^2/4t}\mathcal{A}_i(0,y)$
appearing similar to the coarse-graining
if $t$ is identified with the renormalization scale.
We note that this can be constructed in a gauge-covariant manner.
There are many studies on non-perturbative formulation and nontrivial fixed points of RG
based on the gradient flow~\cite{Kagimura:2015via,Yamamura:2015kva,Aoki:2016ohw,%
Makino:2018rys,Abe:2018zdc,Carosso:2018bmz,Carosso:2018rep,Sonoda:2019ibh,%
Carosso:2019qpb,Nakamura:2019ind,Matsumoto:2020lha,Sonoda:2020vut,Nakamura:2021meh,%
Miyakawa:2021hcx,Abe:2022smm,Sonoda:2022fmk,Schierholz:2022tgb,Tanaka:2022pwt,%
Schierholz:2022wuc,Hasenfratz:2023bok,Miyakawa:2023yob,Haruna:2023spq}
(see also Refs.~\cite{Aoki:2015dla,Aoki:2016env,Aoki:2017bru,Aoki:2017uce,%
Bietenholz:2018agd,Artz:2019bpr}).

In particular, one finds that a dimensionless flowed operator $\mathcal{O}_i(t,x)$
associated with the gauge coupling~$g_i$ provides a renormalization scheme
such that a running coupling~$g_i^2(\mu)_{\mathrm{GF}}$ is non-perturbatively defined as
\begin{equation}
 g_i^2(\mu)_{\mathrm{GF}}
  \equiv \left\langle\mathcal{O}_i(t)\right\rangle_{\sqrt{8t}=1/\mu}
  \sim g_i^2 + O(g_i^4)
\end{equation}
with the renormalization scale~$\mu$ identical to~$1/\sqrt{8t}$.
Furthermore, gearing the flow time to the finite physical box size
in lattice gauge theory, we can compute numerically the running coupling
through a sophisticated finite-size
scaling analysis~\cite{Luscher:1991wu,Luscher:2014kea}.
Based on this method, for instance, the size of the strong coupling $\alpha_s$ in QCD
was determined with high accuracy~\cite{Bruno:2017gxd}.

Suppose that there exists a one-to-one mapping
of the parameter space of \emph{all} coupling constants
into the space spanned by the set $\{\langle\mathcal{O}_i(t)\rangle\}$.
Reference~\cite{Makino:2018rys} addressed the analytical illustration of this RG flow
in the following theories, in which a $2$-dimensional parameter space
$(\langle\mathcal{O}_1(t)\rangle,\langle\mathcal{O}_2(t)\rangle)$
plays an important role:
the two-loop approximation of the $4$-dimensional many-flavor gauge theory
and the large-$N$ limit of the $3$-dimensional ($3$D) $O(N)$ linear sigma model.
Then, we can confirm whether (a combination of) $\langle\mathcal{O}_i(t)\rangle$ is relevant or irrelevant
around an infrared fixed point in the limit $t\to\infty$
by way of illustration. (See also Ref.~\cite{Aoki:2016ohw})

In this paper, we reconsider one example given in~Ref.~\cite{Makino:2018rys},
the $3$D $O(N)$ linear sigma model,
which possesses the Wilson--Fisher fixed point~\cite{Wilson:1971dc}
in the infrared limit.
First, we review the flowed scalar theory
and the construction of flowed operators $\{\mathcal{O}_i\}_{i=1,2}$
under the large~$N$ approximation following Ref.~\cite{Makino:2018rys},
and then compute the critical exponent of the relevant parameter.
Next, we simulate numerically this model with finite~$N$
based on the lattice regularization.
As a completely non-perturbative approach,
we finally depict the RG flow in our parameter space
$(\langle\mathcal{O}_1(t)\rangle, \langle\mathcal{O}_2(t)\rangle)$
and observe the Wilson--Fisher fixed point.

In future,
we hope to get a better understanding of RG in gauge theory
via the gradient flow.
As already mentioned, this new approach to RG is manifestly gauge invariant.
We can simply apply our method to lattice simulations of gauge theory.

\section{Infrared criticality of $O(N)$ sigma model on $3$D continuum spacetime}
\subsection{Gradient flow for scalar fields and its relation with Wilsonian RG}
The $3$D $O(N)$ linear sigma model is defined by the following Euclidean action
\begin{equation}
 S_{\mathrm{E}}=\int d^3x\,
  \left\{
   \frac{1}{2}\left[\partial_\mu\phi_i(x)\right]^2
   + \frac{1}{2}m_0^2\phi_i^2(x)
   + \frac{1}{8N}\lambda_0\left[\phi_i(x)^2\right]^2
  \right\} ,
  \label{eq:action}
\end{equation}
where $i=1$, \dots, $N$.
For the scalar fields $\{\phi_i\}$,
we introduce the flow equation~\cite{Capponi:2015ucc}
\begin{equation}
 \partial_t\varphi_i(t,x) = \partial_\mu^2\varphi_i(t,x) ,
  \qquad
  \varphi_i(t=0,x) = \phi_i(x) .
\end{equation}
It is proved in perturbation theory that,
using the renormalized coupling and the wave function renormalization,
any composite operator of $\varphi_i(t,x)$
is automatically a finite renormalized operator.
The correlation functions of $\varphi_i(t,x)$ can be computed
by substituting
\begin{equation}
 \varphi_i(t,x) = \int d^3y\int\frac{d^3p}{(2\pi)^3} e^{ip(x-y)}e^{-tp^2}\phi_i(x) .
\end{equation}
The ringed field variable~$\mathring{\varphi}_i(t,x)$ defined by
\begin{equation}
 \mathring{\varphi}_i(t,x)
  \equiv \sqrt{\frac{N}{2(2\pi)^{3/2}t^{1/2}\langle\varphi_j(t,x)^2\rangle}}
  \varphi_i(t,x)
  \stackrel{t\to0}{\to} \varphi_i(t,x) + O(1/N)
\end{equation}
is free from the multiplicative renormalization factor.

Assuming the translational invariance for one-point functions,
we see the scaling relation for the flowed operators constructed by~$\mathrm{\varphi}_i$
under~$x\mapsto e^{\xi}x$~\cite{Makino:2018rys}
\begin{equation}
 \left\langle\mathcal{O}_i(e^{2\xi}t)\right\rangle_{\{g_j\}}
  =\left\langle\mathcal{O}_i(t)\right\rangle_{\{g_j(\xi)\}} ,
\end{equation}
up to a nontrivial operator mixing.
The coupling constants, $\{g_i\}$, run along the RG flow parameterized by~$\xi$
as~$\{g_i(\xi)\}$.
In general, on the assumption that the mapping as
\begin{equation}
 g_i(\xi) \mapsto \langle\mathcal{O}_i(t)\rangle = \mathcal{R}_i[\{g_j(\xi)\}]
\end{equation}
is one-to-one,
the set of one-point functions $\{\langle\mathcal{O}_i(t)\rangle\}$
can be regarded as a set of running couplings non-perturbatively defined.

\subsection{Large $N$ solution and the critical exponent}
The solution of the model is well known at the $1/N$ expansion
through the use of the auxiliary field method.
The physical mass scale $M$ is given by the mass-gap equation
\begin{equation}
 M^2 + \frac{\lambda_0}{8\pi}M
  = m_0^2 + \frac{1}{4\pi^2}\lambda_0\Lambda ,
  \label{eq:massgap}
\end{equation}
where $\Lambda$ is the momentum cutoff,
and the renormalized coupling at the renormalization scale~$\mu$ is
\begin{equation}
 \frac{\lambda}{\mu} = \frac{\lambda_0}{\mu}
  \left(1 + \frac{\sqrt{3}}{96}\frac{\lambda_0}{\mu}\right)^{-1} .
\end{equation}
We have the RG equations
\begin{align}
 \beta\left(\frac{\lambda}{\mu}\right)
 \equiv \left(\mu\frac{\partial}{\partial\mu}\right)_0 \frac{\lambda}{\mu}
 &= - \frac{\lambda}{\mu} + \frac{\sqrt{3}}{96}\left(\frac{\lambda}{\mu}\right)^2 ,
 &
 \left(\mu\frac{\partial}{\partial\mu}\right)_0 \frac{M}{\mu}
 &= - \frac{M}{\mu} .
 \label{eq:beta_mu}
\end{align}
The fixed points, that is, zeros of the beta functions~\eqref{eq:beta_mu},
are given at $(\frac{\lambda_*}{\mu},\frac{M_*}{\mu})=(0,0)$
and $(\frac{\lambda_*}{\mu},\frac{M_*}{\mu})=(\frac{96}{\sqrt{3}},0)$.
The critical exponents correspond to the slopes of the beta function $\beta'$
near the fixed points:
\begin{itemize}
 \item At $(\frac{\lambda_*}{\mu},\frac{M_*}{\mu})=(0,0)$,
       $\lambda$ is relevant as $\beta'(\frac{\lambda_*}{\mu})=-1$
       and $M$ is also relevant as $-1$ (Gaussian fixed point).
 \item At $(\frac{\lambda_*}{\mu},\frac{M_*}{\mu})=(\frac{96}{\sqrt{3}},0)$,
       $\lambda$ is irrelevant as $\beta'(\frac{\lambda_*}{\mu})=+1$
       but $M$ is relevant as $-1$ (Wilson--Fisher fixed point).
\end{itemize}

\begin{figure}[t]
 \centering
 \includegraphics[clip,width=0.45\columnwidth]{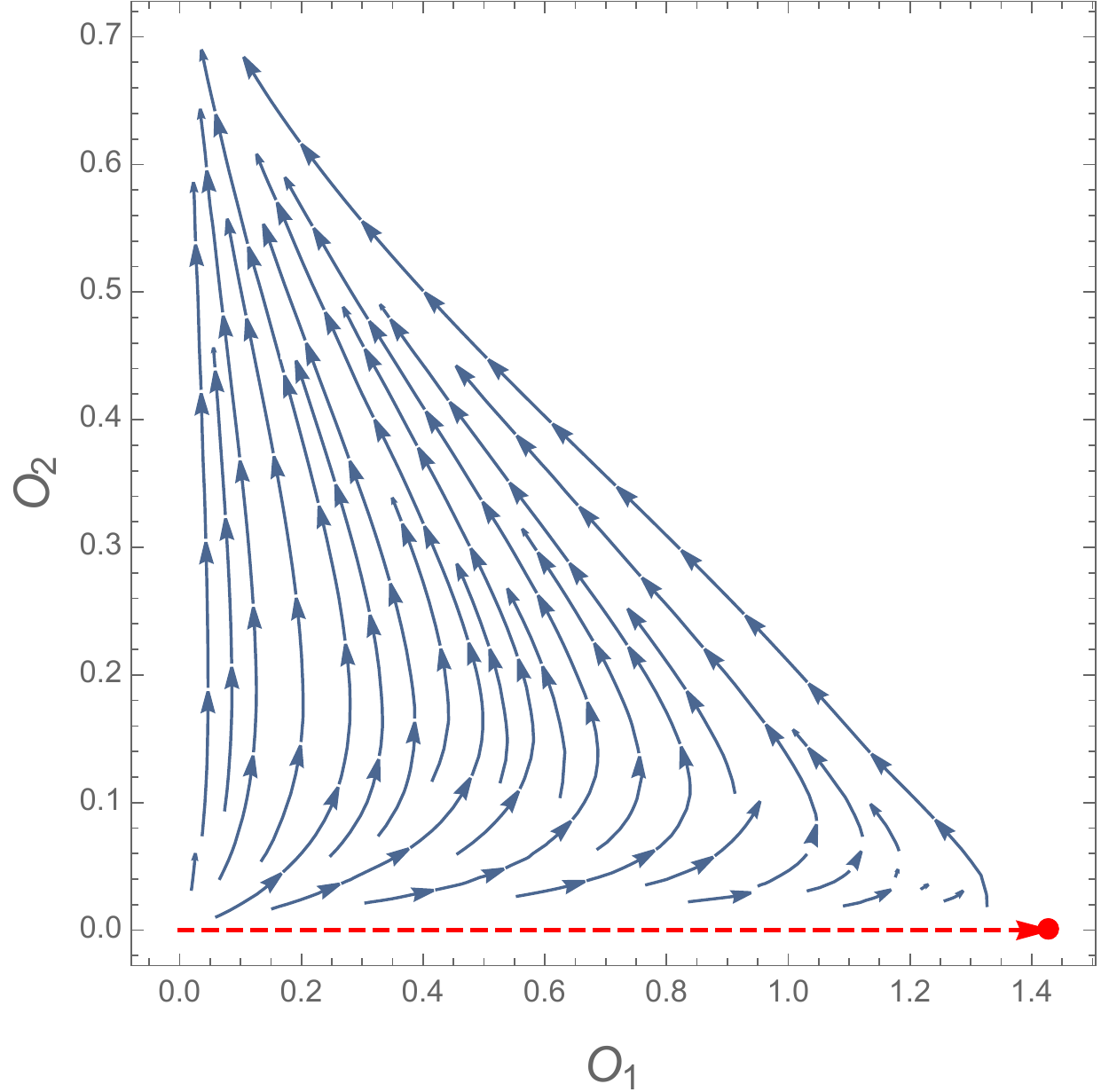}
 \caption{The RG flow of
 $(\langle\mathcal{O}_1(t)\rangle, \langle\mathcal{O}_2(t)\rangle)$
 depicted in Ref.~\cite{Makino:2018rys}}
 \label{fig:makino}
\end{figure}

Instead of the running coupling~$\lambda$ and the mass $M$,
we define corresponding dimensionless operators as follows:
\begin{align}
 \mathcal{O}_1(t,x)
 &\equiv - \frac{4(2\pi)^3}{N}t\left[\mathring{\varphi}(t,x)^2\right]^2
 + N + 2 ,\\
 \mathcal{O}_2(t,x)
 &\equiv \frac{16\pi}{N}t^{3/2}\left[\partial_\mu\mathring{\varphi}_i(t,x)\right]^2
 - \frac{1}{(2\pi)^{1/2}}.
\end{align}
From the analytical computation at the large $N$ approximation,
the asymptotic behaviors are given by
\begin{align}
 \langle\mathcal{O}_1(t)\rangle
 &\stackrel{t\to0}{\to} K\lambda_0 t^{1/2}\\
 &\stackrel{t\to\infty}{\to}
 \begin{cases}
  K'\frac{\lambda_0}{M}
  \left(1+\frac{1}{16\pi}\frac{\lambda_0}{M}\right)^{-1}\frac{1}{M^3t^{3/2}}
  &\text{for $M/\lambda_0>0$},\\
  K_* &\text{for $M/\lambda_0\to0$},
 \end{cases}
 \\
 \langle\mathcal{O}_2(t)\rangle
 &\stackrel{t\to0}{\to} M t^{1/2}\\
 &\stackrel{t\to\infty}{\to} \left(\frac{2}{\pi}\right)^{1/2}
 - \frac{3}{(8\pi)^{1/2}}\frac{1}{M^2t} ,
\end{align}
where
\begin{align}
 K &\cong 0.289432 ,& K' &= \frac{1}{(4\pi)^{3/2}} ,&
 K_* &\cong 1.4259.
\end{align}
Note that $\langle\mathcal{O}_2(t)\rangle\equiv0$ in the limit $M\to0$.
We see the RG flow of the parameter space of~$\langle\mathcal{O}_1(t)\rangle$
and $\langle\mathcal{O}_2(t)\rangle$ arrowed along~$t$
in Fig.~\ref{fig:makino}~\cite{Makino:2018rys}.
The infrared Wilson--Fisher fixed point as depicted by the red point is indicated
at~$(\langle\mathcal{O}_1(t)\rangle,\langle\mathcal{O}_2(t)\rangle)=(K_*,0)$.
This discussion completes the observation given in~Ref.~\cite{Makino:2018rys}.

From now on, let us compute the critical exponent
based on the RG equation with regard to~$\langle\mathcal{O}_1(t)\rangle$
and~$\langle\mathcal{O}_2(t)\rangle$.
We first note that
the theory around the Gaussian fixed point at~$(0,0)$ possesses
the small enough dimensionless parameter $\lambda_0/M$
while that around the Wilson--Fisher fixed point at~$(K_*,0)$ does the large one.
Then, we observe the critical behavior at the Gaussian fixed point
in the $\lambda_0\to0$ limit
\begin{align}
 \langle\mathcal{O}_1(t)\rangle_{\mathrm{G}} &\propto \lambda_0t^{1/2} ,&
 \langle\mathcal{O}_2(t)\rangle_{\mathrm{G}} &\propto M t^{1/2} ,
\end{align}
and therefore we find the same critical exponents as~Eq.~\eqref{eq:beta_mu}
\begin{equation}
 \left(t\frac{d}{dt}\right)\langle\mathcal{O}_i(t)\rangle_{\mathrm{G}}
 = \frac{1}{2}\langle\mathcal{O}_i(t)\rangle_{\mathrm{G}}
 + O(\langle\mathcal{O}_i(t)\rangle_{\mathrm{G}}^2) .
\end{equation}
Here note that the mass dimension of~$t$ is $-2$ while $\lambda$ is $1$.
On the other hand, at the Wilson--Fisher fixed point,
we see the irrelevant behavior in the $M\to0$ limit
\begin{equation}
 \langle\mathcal{O}_1(t)\rangle_{\mathrm{WF}}-K_* \propto \lambda_0^{-1/2}t^{-1/2} ,
 \label{eq:O_largeN_Mzero}
\end{equation}
and the relevant one in the $\lambda_0\to\infty$ limit
\begin{align}
 \langle\mathcal{O}_1(t)\rangle_{\mathrm{WF}}-K_* &\propto M t^{1/2} ,&
 \langle\mathcal{O}_2(t)\rangle_{\mathrm{WF}} &\propto M t^{1/2} .
 \label{eq:O_largeN_lambdainf}
\end{align}
Redefining $\langle\mathcal{O}_2(t)\rangle$ as an appropriate linear combination
of~$(\langle\mathcal{O}_1(t)\rangle-K_*)$ and $\langle\mathcal{O}_2(t)\rangle$,
that is, diagonalizing Eqs.~\eqref{eq:O_largeN_Mzero} and~\eqref{eq:O_largeN_lambdainf},
one finds the RG equations as expected
\begin{align}
 \left(t\frac{d}{dt}\right)\langle\mathcal{O}_1(t)\rangle_{\mathrm{WF}}
 &= -\frac{1}{2}\langle\mathcal{O}_1(t)\rangle_{\mathrm{WF}}
 + O(\langle\mathcal{O}_1(t)\rangle_{\mathrm{WF}}^2) ,\\
 \left(t\frac{d}{dt}\right)\langle\mathcal{O}_2(t)\rangle_{\mathrm{WF}}
 &= \frac{1}{2}\langle\mathcal{O}_2(t)\rangle_{\mathrm{WF}}
 + O(\langle\mathcal{O}_2(t)\rangle_{\mathrm{WF}}^2) .
\end{align}

\section{Lattice simulation of RG flow in $3$D $O(N)$ sigma model}
\begin{figure}[t]
    \centering
    \includegraphics[width=0.5\linewidth]{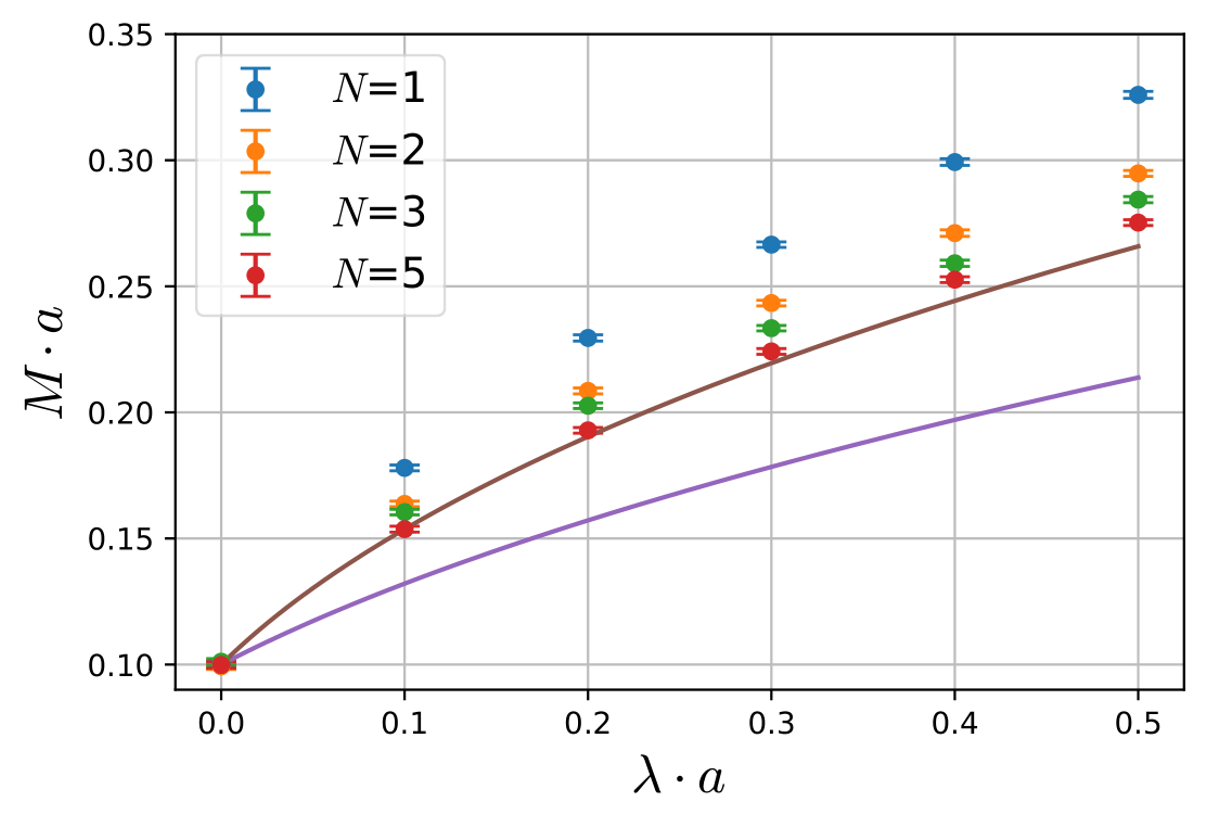}
    \caption{Effective mass, $M$, as a function of the coupling $\lambda$.
    Dependence to $N$ is also shown.
    Two curves are given by the gap equation
    in the large~$N$ approximation in Eq.~\eqref{eq:massgap}
    with $\Lambda=\pi/2 a$ (above orange)
    and $\Lambda=\pi/a$ (below purple).
    Despite the different regularizations,
    the lattice approach seems to be consistent with
    the large~$N$ solution.}
    \label{fig:effmass}
\end{figure}

In this section, we attempt numerical simulations
for the finite-$N$ $O(N)$ sigmal model.
By using the simple symmetric difference instead of the derivative,
the discretized lattice action includes the tunable parameters,
$m_0 a$ and $\lambda_0 a$.
We utilized the overrelaxed heatbath method for configuration generation.
From the computation of the two-point function of~$\phi$, as usual,
we see the coupling dependence of the effective mass in Fig.~\ref{fig:effmass} for~$N=1$, $2$, $3$, $5$,
and also the result of the large~$N$ gap equation.

\begin{figure}[t]
 \centering
 \includegraphics[clip,width=0.48\columnwidth]{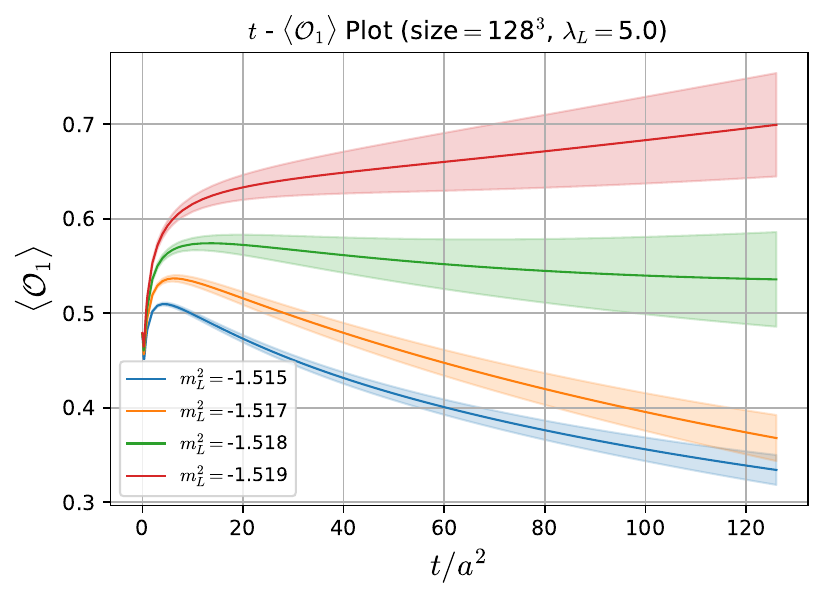}\hspace{1em}
 \includegraphics[clip,width=0.48\columnwidth]{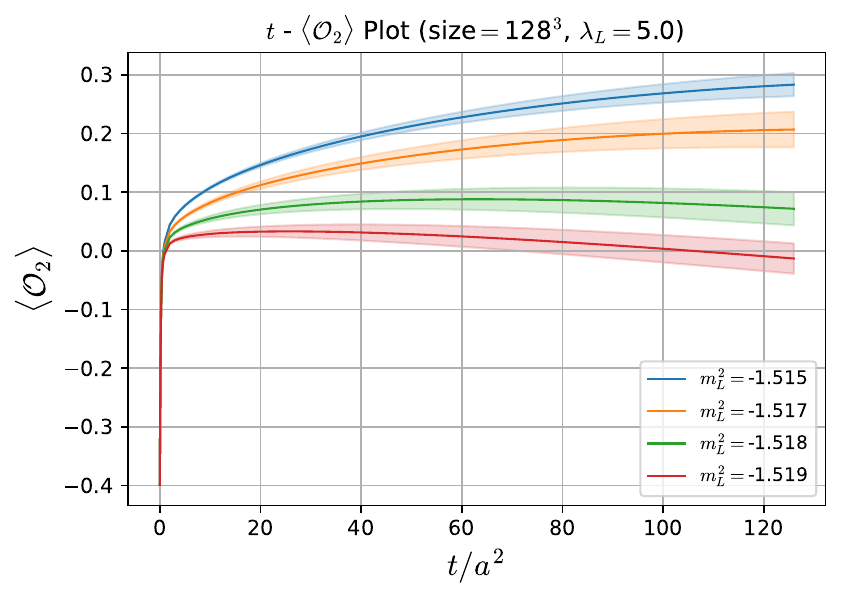}
 \caption{$\langle\mathcal{O}_1(t)\rangle$ (left)
 or $\langle\mathcal{O}_2(t)\rangle$ (right)
 plots as functions of gradient flow time~$t$.
 For a fixed $\lambda_L=\lambda_0 a$, various values of~$m_L=m_0 a$ are shown.
 We see the plateau between $m_L^2=-1.519$ and $-1.518$.}
 \label{fig:O-t}
\end{figure}

\begin{figure}[t]
 \centering
 \includegraphics[clip,width=0.7\columnwidth]{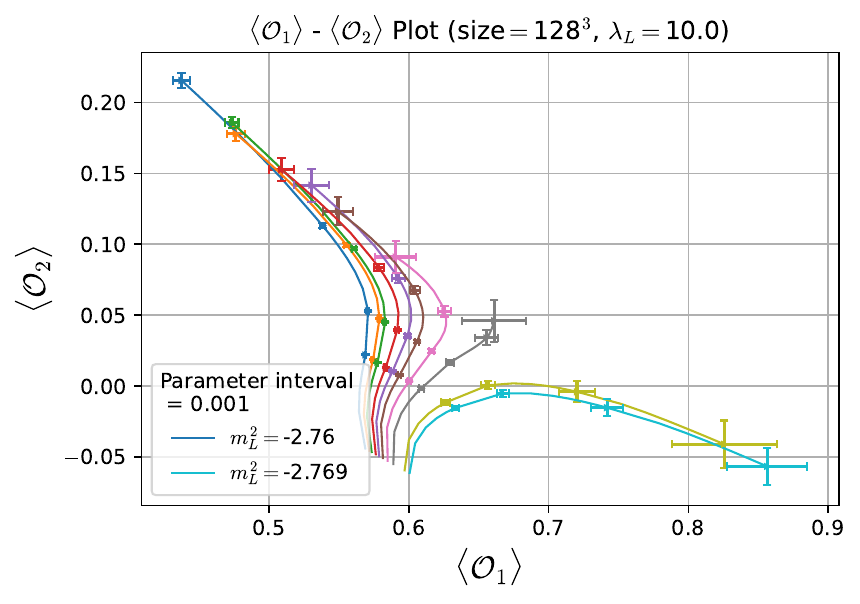}
 \caption{The RG flow of
 $(\langle\mathcal{O}_1(t)\rangle, \langle\mathcal{O}_2(t)\rangle)$ along gradient flow}
 \label{fig:O1O2}
\end{figure}

Now, we again consider the flowed operators,
$\mathcal{O}_1(t)$ and $\mathcal{O}_2(t)$,
via the $4$th order Runge--Kutta method for gradient flow.
To look forward to the critical behavior,
tuning $m_0 a$ so as to make $M$ smaller,
we show the $t$-$\langle\mathcal{O}_i(t)\rangle$ plots
with fixed $\lambda_0 a=5.0$ and~$N=1$ in Fig.~\ref{fig:O-t}
(the lattice size is taken to~$128^3$).
This figure appears to imply the existence of the plateau,
that is, $t$-independent critical couplings for large enough~$t$.
To see this behavior explicitly, Fig.~\ref{fig:O1O2} gives
the $t$ flow of~$\langle\mathcal{O}_1\rangle$ (horizontal axis)
and~$\langle\mathcal{O}_2\rangle$ (vertical axis) simultaneously.
This is just the figure of the RG flow
for the effective couplings,
$\langle\mathcal{O}_1\rangle$ and $\langle\mathcal{O}_2\rangle$.
The left below side means the UV region (small $t$),
while the left above or the right below side is the IR region (large $t$).
In between, the gray curve tends to stop for large $t$, which indicates the existence of the Wilson--Fisher fixed point around there.
The RG trajectories flowing to the left are in the symmetric phase,
while those flowing to the right are in the broken phase.
This is our main result.

Here, from Fig.~\ref{fig:O1O2},
one may ask the following questions:
(i) why the RG trajectories are likely to be crossing each other at large~$t$ (near IR);
(ii) where is the Gaussian fixed point because the point of each trajectory at~$t=0$ (UV) is different.

\begin{figure}[t]
 \centering
 \includegraphics[clip,width=0.485\columnwidth]{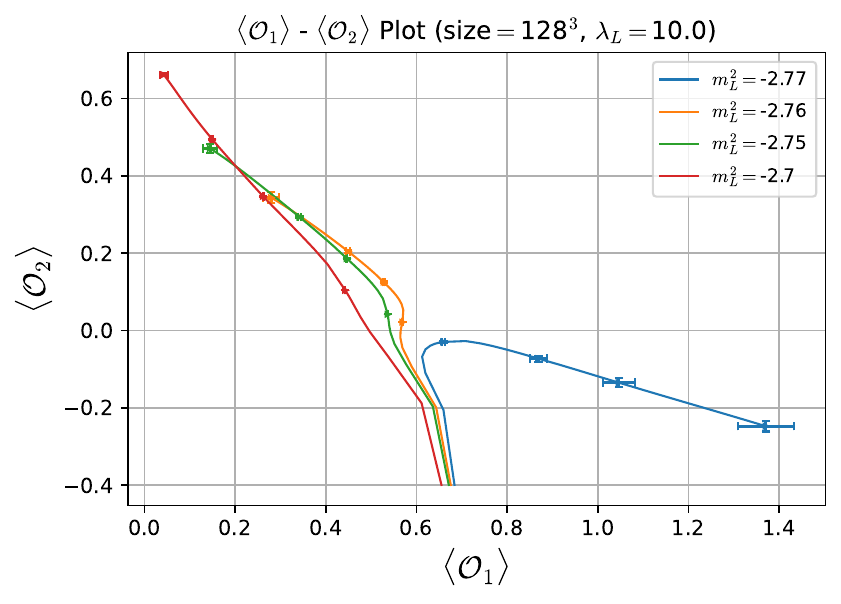}\hspace{0.5em}
 \includegraphics[clip,width=0.485\columnwidth]{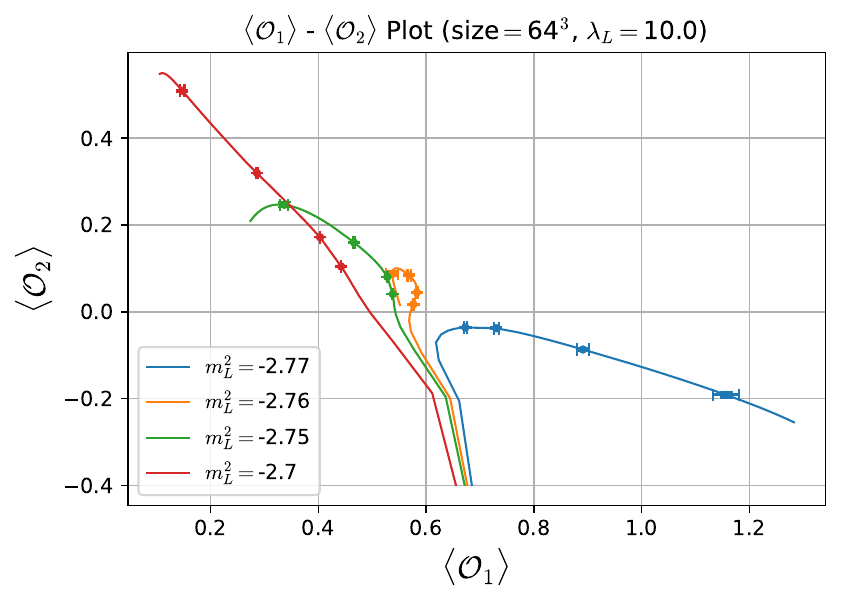}
 \caption{Finite size effect}
 \label{fig:Lsize}
\end{figure}

The first issue happens when the lattice size is too small;
it is a finite-size effect.
To see this, we compare the
$\langle\mathcal{O}_1(t)\rangle$-$\langle\mathcal{O}_2(t)\rangle$ plot
for different lattice sizes.
In Fig.~\ref{fig:Lsize}, for $\lambda_0 a=10.0$,
the left panel is devoted to the larger lattice size $128^3$,
while the right panel is to the smaller lattice size $64^3$.
Each trajectory then suffers from more severe intersections
if the lattice size is smaller.
In fact, for such a large flow time,
$\langle\mathcal{O}_i(t)\rangle$ is oversmeared
such that the diffusion length is comparable to the lattice size.

\begin{figure}[t]
 \centering
 \includegraphics[clip,width=0.48\columnwidth]{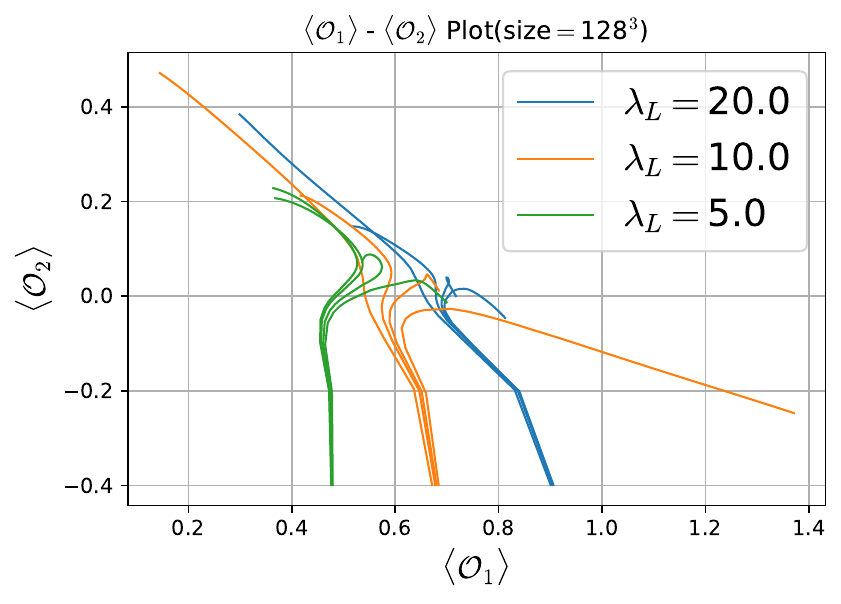}
 \caption{Lattice model details at UV}
 \label{fig:lambda}
\end{figure}

For the second point, that is, the UV behavior,
we can simply say that the system is not taken to the continuum limit,
and then is just a finite lattice model.
Actually, for different values of~$\lambda$,
Fig.~\ref{fig:lambda} shows that small $\lambda$ (and small $M$)
makes $\langle\mathcal{O}_1\rangle$ small.
For the continuum limit, we expect the Gaussian fixed point
with $\lambda$ and $M$ near zero.
Also, the critical behavior for each lattice model should be the same
near the Wilson--Fisher fixed point.
We can consider the following two situations:
\begin{itemize}
    \item For small $\lambda$, the system is close to the Gaussian
    and far from the Wilson--Fisher fixed point.
    We need to have a sufficiently larger lattice size
    and try to reduce numerical errors for solving the gradient flow.
    \item For large $\lambda$, the system is far from the Gaussian
    but hopefully close to the Wilson--Fisher fixed point.
    More computational costs when generating configurations
    is predictable for strongly coupled theories.
\end{itemize}

\subsection*{Acknowledgements}
This work was partially supported by Japan Society for the Promotion of Science (JSPS)
Grant-in-Aid for Scientific Research Grant Numbers
JP19H05598, JP22K03619, JP23H04507, JP24K07049 (M.K.),
JP21J30003, JP22KJ2096 (O.M.),
and JP23K03418 (H.S.).
O.M.\ acknowledges the RIKEN Special Postdoctoral Researcher Program.

\bibliographystyle{JHEP}
\bibliography{ref}

\providecommand{\href}[2]{#2}\begingroup\raggedright\begin{thebibliography}{10}

\bibitem{Wilson:1971bg}
K.G.~Wilson, \emph{{Renormalization group and critical phenomena. 1. Renormalization group and the Kadanoff scaling picture}}, \href{https://doi.org/10.1103/PhysRevB.4.3174}{\emph{Phys. Rev. B} {\bfseries 4} (1971) 3174}.

\bibitem{Wilson:1971dh}
K.G.~Wilson, \emph{{Renormalization group and critical phenomena. 2. Phase space cell analysis of critical behavior}}, \href{https://doi.org/10.1103/PhysRevB.4.3184}{\emph{Phys. Rev. B} {\bfseries 4} (1971) 3184}.

\bibitem{Kadanoff:1966wm}
L.P.~Kadanoff, \emph{{Scaling laws for Ising models near T(c)}}, \href{https://doi.org/10.1103/PhysicsPhysiqueFizika.2.263}{\emph{Physics Physique Fizika} {\bfseries 2} (1966) 263}.

\bibitem{Mack:1969rr}
G.~Mack and A.~Salam, \emph{{Finite component field representations of the conformal group}}, \href{https://doi.org/10.1016/0003-4916(69)90278-4}{\emph{Annals Phys.} {\bfseries 53} (1969) 174}.

\bibitem{Polyakov:1970xd}
A.M.~Polyakov, \emph{{Conformal symmetry of critical fluctuations}}, {\emph{JETP Lett.} {\bfseries 12} (1970) 381}.

\bibitem{Wilson:1973jj}
K.G.~Wilson and J.B.~Kogut, \emph{{The Renormalization group and the epsilon expansion}}, \href{https://doi.org/10.1016/0370-1573(74)90023-4}{\emph{Phys. Rept.} {\bfseries 12} (1974) 75}.

\bibitem{Zinn-Justin:2002ecy}
J.~Zinn-Justin, \emph{{Quantum Field Theory and Critical Phenomena}}, Oxford University Press (06, 2002), \href{https://doi.org/10.1093/acprof:oso/9780198509233.001.0001}{10.1093/acprof:oso/9780198509233.001.0001}.

\bibitem{Sonoda:2007dj}
H.~Sonoda, \emph{{On the construction of QED using ERG}}, \href{https://doi.org/10.1088/1751-8113/40/31/034}{\emph{J. Phys. A} {\bfseries 40} (2007) 9675} [\href{https://arxiv.org/abs/hep-th/0703167}{{\ttfamily hep-th/0703167}}].

\bibitem{Igarashi:2009tj}
Y.~Igarashi, K.~Itoh and H.~Sonoda, \emph{{Realization of Symmetry in the ERG Approach to Quantum Field Theory}}, \href{https://doi.org/10.1143/PTPS.181.1}{\emph{Prog. Theor. Phys. Suppl.} {\bfseries 181} (2010) 1} [\href{https://arxiv.org/abs/0909.0327}{{\ttfamily 0909.0327}}].

\bibitem{Luscher:2013vga}
M.~L\"uscher, \emph{{Future applications of the Yang-Mills gradient flow in lattice QCD}}, \href{https://doi.org/10.22323/1.187.0016}{\emph{PoS} {\bfseries LATTICE2013} (2014) 016} [\href{https://arxiv.org/abs/1308.5598}{{\ttfamily 1308.5598}}].

\bibitem{Narayanan:2006rf}
R.~Narayanan and H.~Neuberger, \emph{{Infinite N phase transitions in continuum Wilson loop operators}}, \href{https://doi.org/10.1088/1126-6708/2006/03/064}{\emph{JHEP} {\bfseries 03} (2006) 064} [\href{https://arxiv.org/abs/hep-th/0601210}{{\ttfamily hep-th/0601210}}].

\bibitem{Luscher:2009eq}
M.~L\"uscher, \emph{{Trivializing maps, the Wilson flow and the HMC algorithm}}, \href{https://doi.org/10.1007/s00220-009-0953-7}{\emph{Commun. Math. Phys.} {\bfseries 293} (2010) 899} [\href{https://arxiv.org/abs/0907.5491}{{\ttfamily 0907.5491}}].

\bibitem{Luscher:2010iy}
M.~L\"uscher, \emph{{Properties and uses of the Wilson flow in lattice QCD}}, \href{https://doi.org/10.1007/JHEP08(2010)071}{\emph{JHEP} {\bfseries 08} (2010) 071} [\href{https://arxiv.org/abs/1006.4518}{{\ttfamily 1006.4518}}].

\bibitem{Luscher:2011bx}
M.~L\"uscher and P.~Weisz, \emph{{Perturbative analysis of the gradient flow in non-abelian gauge theories}}, \href{https://doi.org/10.1007/JHEP02(2011)051}{\emph{JHEP} {\bfseries 02} (2011) 051} [\href{https://arxiv.org/abs/1101.0963}{{\ttfamily 1101.0963}}].

\bibitem{Kagimura:2015via}
A.~Kagimura, A.~Tomiya and R.~Yamamura, \emph{{Effective lattice action for the configurations smeared by the Wilson flow}},  \href{https://arxiv.org/abs/1508.04986}{{\ttfamily 1508.04986}}.

\bibitem{Yamamura:2015kva}
R.~Yamamura, \emph{{The Yang\textemdash{}Mills gradient flow and lattice effective action}}, \href{https://doi.org/10.1093/ptep/ptw097}{\emph{PTEP} {\bfseries 2016} (2016) 073B10} [\href{https://arxiv.org/abs/1510.08208}{{\ttfamily 1510.08208}}].

\bibitem{Aoki:2016ohw}
S.~Aoki, J.~Balog, T.~Onogi and P.~Weisz, \emph{{Flow equation for the large $N$ scalar model and induced geometries}}, \href{https://doi.org/10.1093/ptep/ptw106}{\emph{PTEP} {\bfseries 2016} (2016) 083B04} [\href{https://arxiv.org/abs/1605.02413}{{\ttfamily 1605.02413}}].

\bibitem{Makino:2018rys}
H.~Makino, O.~Morikawa and H.~Suzuki, \emph{{Gradient flow and the Wilsonian renormalization group flow}}, \href{https://doi.org/10.1093/ptep/pty050}{\emph{PTEP} {\bfseries 2018} (2018) 053B02} [\href{https://arxiv.org/abs/1802.07897}{{\ttfamily 1802.07897}}].

\bibitem{Abe:2018zdc}
Y.~Abe and M.~Fukuma, \emph{{Gradient flow and the renormalization group}}, \href{https://doi.org/10.1093/ptep/pty081}{\emph{PTEP} {\bfseries 2018} (2018) 083B02} [\href{https://arxiv.org/abs/1805.12094}{{\ttfamily 1805.12094}}].

\bibitem{Carosso:2018bmz}
A.~Carosso, A.~Hasenfratz and E.T.~Neil, \emph{{Nonperturbative Renormalization of Operators in Near-Conformal Systems Using Gradient Flows}}, \href{https://doi.org/10.1103/PhysRevLett.121.201601}{\emph{Phys. Rev. Lett.} {\bfseries 121} (2018) 201601} [\href{https://arxiv.org/abs/1806.01385}{{\ttfamily 1806.01385}}].

\bibitem{Carosso:2018rep}
A.~Carosso, A.~Hasenfratz and E.T.~Neil, \emph{{Renormalization group properties of scalar field theories using gradient flow}}, \href{https://doi.org/10.22323/1.334.0248}{\emph{PoS} {\bfseries LATTICE2018} (2018) 248} [\href{https://arxiv.org/abs/1811.03182}{{\ttfamily 1811.03182}}].

\bibitem{Sonoda:2019ibh}
H.~Sonoda and H.~Suzuki, \emph{{Derivation of a gradient flow from the exact renormalization group}}, \href{https://doi.org/10.1093/ptep/ptz020}{\emph{PTEP} {\bfseries 2019} (2019) 033B05} [\href{https://arxiv.org/abs/1901.05169}{{\ttfamily 1901.05169}}].

\bibitem{Carosso:2019qpb}
A.~Carosso, \emph{{Stochastic Renormalization Group and Gradient Flow}}, \href{https://doi.org/10.1007/JHEP01(2020)172}{\emph{JHEP} {\bfseries 01} (2020) 172} [\href{https://arxiv.org/abs/1904.13057}{{\ttfamily 1904.13057}}].

\bibitem{Nakamura:2019ind}
Y.~Nakamura and G.~Schierholz, \emph{{Does confinement imply CP invariance of the strong interactions?}}, \href{https://doi.org/10.22323/1.363.0172}{\emph{PoS} {\bfseries LATTICE2019} (2019) 172} [\href{https://arxiv.org/abs/1912.03941}{{\ttfamily 1912.03941}}].

\bibitem{Matsumoto:2020lha}
M.~Matsumoto, G.~Tanaka and A.~Tsuchiya, \emph{{The renormalization group and the diffusion equation}}, \href{https://doi.org/10.1093/ptep/ptaa175}{\emph{PTEP} {\bfseries 2021} (2021) 023B02} [\href{https://arxiv.org/abs/2011.14687}{{\ttfamily 2011.14687}}].

\bibitem{Sonoda:2020vut}
H.~Sonoda and H.~Suzuki, \emph{{Gradient flow exact renormalization group}}, \href{https://doi.org/10.1093/ptep/ptab006}{\emph{PTEP} {\bfseries 2021} (2021) 023B05} [\href{https://arxiv.org/abs/2012.03568}{{\ttfamily 2012.03568}}].

\bibitem{Nakamura:2021meh}
Y.~Nakamura and G.~Schierholz, \emph{{The strong CP problem solved by itself due to long-distance vacuum effects}}, \href{https://doi.org/10.1016/j.nuclphysb.2022.116063}{\emph{Nucl. Phys. B} {\bfseries 986} (2023) 116063} [\href{https://arxiv.org/abs/2106.11369}{{\ttfamily 2106.11369}}].

\bibitem{Miyakawa:2021hcx}
Y.~Miyakawa and H.~Suzuki, \emph{{Gradient flow exact renormalization group: Inclusion of fermion fields}}, \href{https://doi.org/10.1093/ptep/ptab100}{\emph{PTEP} {\bfseries 2021} (2021) 083B04} [\href{https://arxiv.org/abs/2106.11142}{{\ttfamily 2106.11142}}].

\bibitem{Abe:2022smm}
Y.~Abe, Y.~Hamada and J.~Haruna, \emph{{Fixed point structure of the gradient flow exact renormalization group for scalar field theories}}, \href{https://doi.org/10.1093/ptep/ptac021}{\emph{PTEP} {\bfseries 2022} (2022) 033B03} [\href{https://arxiv.org/abs/2201.04111}{{\ttfamily 2201.04111}}].

\bibitem{Sonoda:2022fmk}
H.~Sonoda and H.~Suzuki, \emph{{One-particle irreducible Wilson action in the gradient flow exact renormalization group formalism}}, \href{https://doi.org/10.1093/ptep/ptac047}{\emph{PTEP} {\bfseries 2022} (2022) 053B01} [\href{https://arxiv.org/abs/2201.04448}{{\ttfamily 2201.04448}}].

\bibitem{Schierholz:2022tgb}
G.~Schierholz, \emph{{Strong CP problem, electric dipole moment, and fate of the axion}}, \href{https://doi.org/10.21468/SciPostPhysProc.6.011}{\emph{SciPost Phys. Proc.} {\bfseries 6} (2022) 011} [\href{https://arxiv.org/abs/2201.12875}{{\ttfamily 2201.12875}}].

\bibitem{Tanaka:2022pwt}
G.~Tanaka and A.~Tsuchiya, \emph{{Higher-derivative extension of the functional renormalization group}}, \href{https://doi.org/10.1093/ptep/ptac080}{\emph{PTEP} {\bfseries 2022} (2022) 063B02} [\href{https://arxiv.org/abs/2203.07009}{{\ttfamily 2203.07009}}].

\bibitem{Schierholz:2022wuc}
G.~Schierholz, \emph{{Dynamical solution of the strong CP problem within QCD?}}, \href{https://doi.org/10.1051/epjconf/202227401009}{\emph{EPJ Web Conf.} {\bfseries 274} (2022) 01009} [\href{https://arxiv.org/abs/2212.05485}{{\ttfamily 2212.05485}}].

\bibitem{Hasenfratz:2023bok}
A.~Hasenfratz, C.T.~Peterson, J.~van Sickle and O.~Witzel, \emph{{\ensuremath{\Lambda} parameter of the SU(3) Yang-Mills theory from the continuous \ensuremath{\beta} function}}, \href{https://doi.org/10.1103/PhysRevD.108.014502}{\emph{Phys. Rev. D} {\bfseries 108} (2023) 014502} [\href{https://arxiv.org/abs/2303.00704}{{\ttfamily 2303.00704}}].

\bibitem{Miyakawa:2023yob}
Y.~Miyakawa, H.~Sonoda and H.~Suzuki, \emph{{Chiral anomaly as a composite operator in the gradient flow exact renormalization group formalism}}, \href{https://doi.org/10.1093/ptep/ptad074}{\emph{PTEP} {\bfseries 2023} (2023) 063B03} [\href{https://arxiv.org/abs/2304.14753}{{\ttfamily 2304.14753}}].

\bibitem{Haruna:2023spq}
J.~Haruna and M.~Yamada, \emph{{Gradient Flow Exact Renormalization Group for Scalar Quantum Electrodynamics}},  \href{https://arxiv.org/abs/2312.15673}{{\ttfamily 2312.15673}}.

\bibitem{Aoki:2015dla}
S.~Aoki, K.~Kikuchi and T.~Onogi, \emph{{Geometries from field theories}}, \href{https://doi.org/10.1093/ptep/ptv131}{\emph{PTEP} {\bfseries 2015} (2015) 101B01} [\href{https://arxiv.org/abs/1505.00131}{{\ttfamily 1505.00131}}].

\bibitem{Aoki:2016env}
S.~Aoki, J.~Balog, T.~Onogi and P.~Weisz, \emph{{Flow equation for the scalar model in the large $N$ expansion and its applications}}, \href{https://doi.org/10.1093/ptep/ptx025}{\emph{PTEP} {\bfseries 2017} (2017) 043B01} [\href{https://arxiv.org/abs/1701.00046}{{\ttfamily 1701.00046}}].

\bibitem{Aoki:2017bru}
S.~Aoki and S.~Yokoyama, \emph{{Flow equation, conformal symmetry, and anti-de Sitter geometry}}, \href{https://doi.org/10.1093/ptep/pty013}{\emph{PTEP} {\bfseries 2018} (2018) 031B01} [\href{https://arxiv.org/abs/1707.03982}{{\ttfamily 1707.03982}}].

\bibitem{Aoki:2017uce}
S.~Aoki and S.~Yokoyama, \emph{{AdS geometry from CFT on a general conformally flat manifold}}, \href{https://doi.org/10.1016/j.nuclphysb.2018.06.004}{\emph{Nucl. Phys. B} {\bfseries 933} (2018) 262} [\href{https://arxiv.org/abs/1709.07281}{{\ttfamily 1709.07281}}].

\bibitem{Bietenholz:2018agd}
W.~Bietenholz, P.~de~Forcrand, U.~Gerber, H.~Mej\'\i{}a-D\'\i{}az and I.O.~Sandoval, \emph{{Topological Susceptibility of the 2d O(3) Model under Gradient Flow}}, \href{https://doi.org/10.1103/PhysRevD.98.114501}{\emph{Phys. Rev. D} {\bfseries 98} (2018) 114501} [\href{https://arxiv.org/abs/1808.08129}{{\ttfamily 1808.08129}}].

\bibitem{Artz:2019bpr}
J.~Artz, R.V.~Harlander, F.~Lange, T.~Neumann and M.~Prausa, \emph{{Results and techniques for higher order calculations within the gradient-flow formalism}}, \href{https://doi.org/10.1007/JHEP06(2019)121}{\emph{JHEP} {\bfseries 06} (2019) 121} [\href{https://arxiv.org/abs/1905.00882}{{\ttfamily 1905.00882}}].

\bibitem{Luscher:1991wu}
M.~L\"uscher, P.~Weisz and U.~Wolff, \emph{{A Numerical method to compute the running coupling in asymptotically free theories}}, \href{https://doi.org/10.1016/0550-3213(91)90298-C}{\emph{Nucl. Phys. B} {\bfseries 359} (1991) 221}.

\bibitem{Luscher:2014kea}
M.~L\"uscher, \emph{{Step scaling and the Yang-Mills gradient flow}}, \href{https://doi.org/10.1007/JHEP06(2014)105}{\emph{JHEP} {\bfseries 06} (2014) 105} [\href{https://arxiv.org/abs/1404.5930}{{\ttfamily 1404.5930}}].

\bibitem{Bruno:2017gxd}
{\scshape ALPHA} collaboration, \emph{{QCD Coupling from a Nonperturbative Determination of the Three-Flavor $\Lambda$ Parameter}}, \href{https://doi.org/10.1103/PhysRevLett.119.102001}{\emph{Phys. Rev. Lett.} {\bfseries 119} (2017) 102001} [\href{https://arxiv.org/abs/1706.03821}{{\ttfamily 1706.03821}}].

\bibitem{Wilson:1971dc}
K.G.~Wilson and M.E.~Fisher, \emph{{Critical exponents in 3.99 dimensions}}, \href{https://doi.org/10.1103/PhysRevLett.28.240}{\emph{Phys. Rev. Lett.} {\bfseries 28} (1972) 240}.

\bibitem{Capponi:2015ucc}
F.~Capponi, A.~Rago, L.~Del~Debbio, S.~Ehret and R.~Pellegrini, \emph{{Renormalisation of the energy-momentum tensor in scalar field theory using the Wilson flow}}, \href{https://doi.org/10.22323/1.251.0306}{\emph{PoS} {\bfseries LATTICE2015} (2016) 306} [\href{https://arxiv.org/abs/1512.02851}{{\ttfamily 1512.02851}}].

\end{thebibliography}\endgroup

\end{document}